\documentclass[conference]{IEEEtran}
\IEEEoverridecommandlockouts
\usepackage{cite}
\usepackage{amsmath,amssymb,amsfonts}
\usepackage{algorithm}

\usepackage{listings}
\usepackage{algorithm}
\usepackage{graphicx}
\usepackage{textcomp}
\usepackage{algpseudocode} 
\usepackage{float} 
\usepackage{xcolor}

\usepackage{enumitem}
\usepackage{tabularx}

\begin{document}

\title{IndexMAC: A Custom RISC-V Vector Instruction to Accelerate Structured-Sparse Matrix Multiplications}
\author{
\IEEEauthorblockN{V. Titopoulos, K. Alexandridis, C. Peltekis}
\IEEEauthorblockA{\footnotesize Electrical and Computer Engineering\\ 
Democritus University of Thrace, Greece\thanks{This work was supported by a research grant from Codasip, a provider of customizable RISC-V IP and Codasip Studio design toolset, and its University Program to Democritus Univ. of Thrace for ``RISCV vector processor design''.}}
\and
\IEEEauthorblockN{C. Nicopoulos}
\IEEEauthorblockA{\footnotesize Electrical and Computer Engineering\\  University of Cyprus, Cyprus}
\and
\IEEEauthorblockN{G. Dimitrakopoulos}
\IEEEauthorblockA{\footnotesize Electrical and Computer Engineering\\ 
Democritus University of Thrace, Greece}}

\maketitle
\begin{abstract}
Structured sparsity has been proposed as an efficient way to prune the complexity of modern Machine Learning (ML) applications and to simplify the handling of sparse data in hardware. The acceleration of ML models -- for both training and inference -- relies primarily on equivalent matrix multiplications that can be executed efficiently on vector processors or custom matrix engines. The goal of this work is to incorporate the simplicity of structured sparsity into vector execution, thereby accelerating the corresponding matrix multiplications. Toward this objective, a new vector index-multiply-accumulate instruction is proposed, which enables the implementation of low-cost indirect reads from the vector register file. This reduces unnecessary memory traffic and increases data locality. The proposed new instruction was integrated in a decoupled RISC-V vector processor with negligible hardware cost. Extensive evaluation demonstrates significant speedups of 1.80$\times$--2.14$\times$, as compared to state-of-the-art vectorized kernels, when executing layers of varying sparsity from  state-of-the-art Convolutional Neural Networks (CNNs).
\end{abstract}


\section{Introduction}

The computation of Machine Learning (ML) models relies primarily on equivalent matrix multiplications that can be efficiently executed by vector processors~\cite{vector-processors}, or specialized matrix engines built on top of systolic arrays~\cite{vegeta}. To reduce memory storage and computation cost, the weights of ML models are often pruned, thereby leading to sparse models~\cite{hoefler2021sparsity}. The derived zero weights are not stored and the corresponding computation is skipped. 

\begin{figure}
    \centering
    \includegraphics[width=0.98\columnwidth]{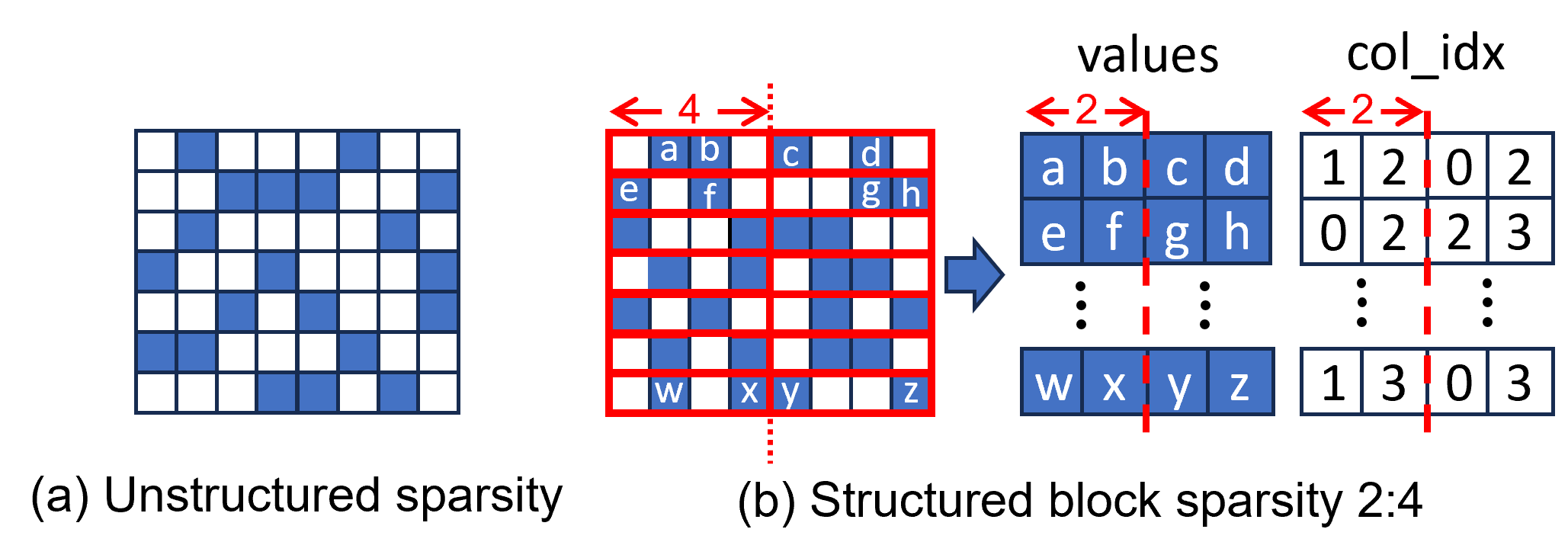}
    \caption{Example of (a) unstructured sparsity; and (b) structured block sparsity of 2:4 (i.e., up to 2 non-zero elements in every 4 consecutive elements) and their respective representation. Each blue square represents a non-zero element.}
    \label{f:unstructered-block-sparse}
\end{figure}

The achieved sparsity can either be \textit{unstructured}~\cite{rigl}, or \textit{structured}~\cite{nvidia-block-sparse,learning-n-m}. In unstructured sparsity, there is no constraint on the locations of the zeros, as shown in Fig.~\ref{f:unstructered-block-sparse}(a). In this case, together with the non-zero elements, multiple indexes are also required to identify the original position of each non-zero element. 

On the contrary, in structured sparsity, there is an upper limit on the number of non-zero elements that may be present within a block of consecutive elements. For instance, in Fig.~\ref{f:unstructered-block-sparse}(b), for every 4 elements in each row, there are up to two non-zero elements. Such block-based structure simplifies both the indexing required to identify the position of each non-zero element inside each block, and the hardware needed to operate on such sparse data. 
In most practical applications~\cite{nvidia-block-sparse,s2ta}, blocks are small and $N$:$M$ sparsity patterns of 1:2, 1:4 or 2:4 are supported, where each block of $M$ elements may contain up to $N$ non-zero elements. 

Structured sparsity promises high performance and low storage overhead. However, in certain cases, it may also lead to less accurate ML models, due to the constrained sparsification process~\cite{sparse-tensor-core, hoefler2021sparsity}. If sparsification involves model retraining, the possible accuracy loss is ameliorated by allowing the model to adapt to the removal of certain weights. 

In this work, our goal is to incorporate the simplicity of structured sparsity into vector execution at the minimum hardware cost. Prior work on vector processors exploits un-structured sparsity to achieve significant speedups. Various sparse vector matrix multiplication algorithms have been developed~\cite{spa, esc}, and recently extended~\cite{generic-spgemm, upc-spgemm}, with the goal being to improve performance by efficiently handling unpredictable sparsity patterns. Besides such software-only approaches, VIA~\cite{via} adds a scratchpad memory and new custom vector instructions to deal with the complexity of 
unstructured sparse data.

Aiming to avoid the substantial hardware overhead incurred in exploiting \textit{unstructured} sparsity, this work makes the case that \emph{structured} sparsity, which appears frequently in state-of-the-art CNN applications~\cite{nvidia-block-sparse}, can be exploited very effectively -- and with negligible hardware cost -- in vector processors to achieve high-performance sparse$\times$dense matrix ($A\times B$) multiplications. 

To achieve this goal we follow a three-fold approach.
First, we adopt a vectorized kernel of the row-based matrix multiplication algorithm that yields good performance under both dense and sparse workloads~\cite{generic-spgemm, matraptor, arm-patent}. Other widely-applicable vectorized algorithms for sparse data, such as SPA~\cite{spa} and its variants~\cite{generic-spgemm, upc-spgemm}, are avoided, since they target highly sparse matrices and do not perform as well with structured sparse data. Second, tiles of matrix $B$, which is treated as dense, are loaded in the vector register file to increase locality. 
Pre-loaded tiles are replaced only when they has been fully used for matrix multiplication and are no longer needed. 
The third and most important step is the introduction of a new vector index-multiply-accumulate instruction that enables the implementation of low-cost indirect reads from the portion of the vector register file where tiles of matrix $B$ reside. This new instruction exploits the structured-sparse format of matrix $A$ to reduce effectively the number of instructions executed per iteration.

Overall, the contributions of this work can be summarized as follows:

\begin{itemize}

\item A state-of-the-art vectorized sparse matrix multiplication~\cite{generic-spgemm, arm-patent} is re-organized to enable the handling of structured sparse data in long-vector ISAs, such as RISC-V. Structured sparsity of one matrix allows us to keep tiles of the other dense matrix 
in the vector register file, thus improving data locality and simplifying loop unrolling~\cite{arm-sve-mm, micro-kernels}.

\item The reformed vectorized matrix multiplication is accelerated with a new index-multiply-accumulate ({\tt vindexmac}) instruction, which enables the implementation of low-cost indirect reads from the vector register file. This eliminates unnecessary memory traffic often encountered in sparse matrix multiplication algorithms and reduces the number of instructions per iteration.

\item The proposed vector instruction was integrated in a high-end decoupled RISC-V vector processor at negligible hardware cost. Extensive evaluation in Gem5~\cite{gem5-orig, gem5-2020} of the vector processor attached to a superscalar out-of-order core demonstrates significant speedups of 1.80$\times$--2.14$\times$, as compared to a state-of-the-art vectorized kernel, when executing state-of-the-art Convolutional Neural Networks (CNNs) pruned for structured sparsity.
\end{itemize}

The rest of the paper is organized as follows: Section~\ref{s:vectorized_sparse} presents the formulation of a state-of-the-art vectorized algorithm for sparse matrix multiplications with structured-sparse data. Section~\ref{s:optimization} introduces the proposed vector index-multiply-accumulate instruction and its hardware implementation. Experimental results are presented in Section~\ref{s:eval} and conclusions are drawn in Section~\ref{s:conclusions}.

\section{Vector Sparse$\times$ Dense Matrix Multiplication}
\label{s:vectorized_sparse}

Vectorized matrix multiplications with sparse data can be implemented with many approaches~\cite{spa, generic-spgemm, upc-spgemm}. The row-wise approach~\cite{matraptor, arm-patent, mm-gp-simd}, also known as Gustavson’s algorithm~\cite{gustavson}, has been shown to be highly effective in computing the matrix product $A\times B$, and it is a better fit to the targeted structured sparsity context. Other approaches~\cite{spa,generic-spgemm,upc-spgemm} target extremely high sparseness and are not as effective with structured sparsity that exhibits modest sparsity.

\begin{figure}[t]
\centering
\includegraphics[width=0.8\columnwidth]{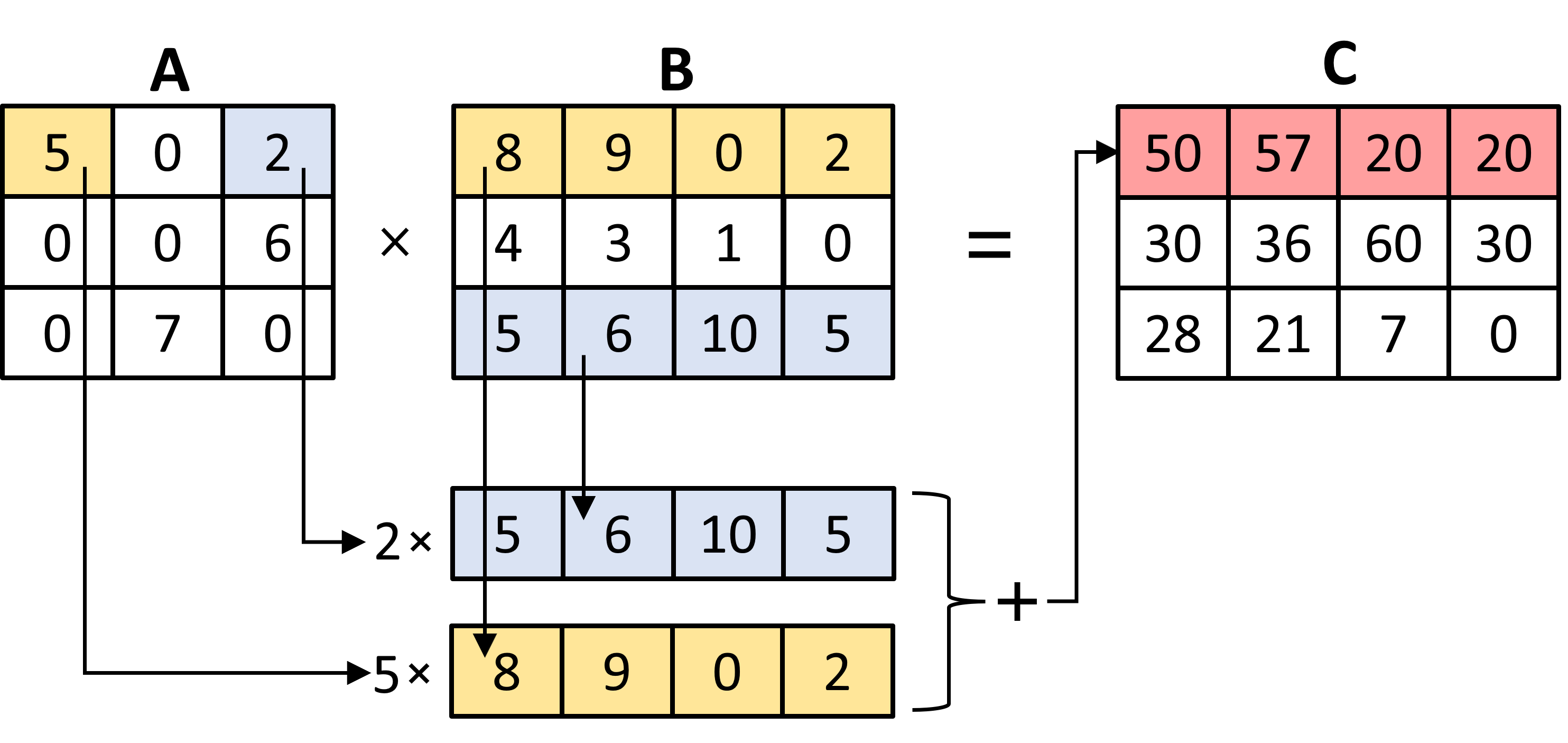}
\caption{Row-wise matrix multiplication to compute output row $C[0,:]$.}
\label{f:row-wise_multiplication}
\end{figure}

Matrix $A$ is sparse and assumed to follow a structured-sparsity template, and, without loss of generality, matrix $B$ is considered to be dense. Algorithmically, all the non-zero elements in a single row of matrix $A$ should be multiplied in parallel with the corresponding rows of matrix $B$, where the row index of matrix $B$ is determined by the column index of the non-zero value in matrix $A$. The product of the multiplication is produced row-by-row, as follows: 
\begin{equation}
C[i,:] = \sum_{k} A[i,k] B[k,:]
\end{equation}
Fig.~\ref{f:row-wise_multiplication} illustrates a simple example of how row $0$ of the result matrix $C$ is produced. Row-wise matrix multiplication can be easily vectorized, since each element of $A$ is multiplied with \textit{all} the elements of a row of matrix $B$. This yields a vector of partial results for a row in the result matrix $C$.

Algorithm~\ref{a:vectorized_row-wise} depicts the vectorized form of the row-wise matrix-multiplication algorithm, \emph{ignoring for clarity} any sparseness in matrix $A$ and assuming an arbitrarily large vector length.  
In line 3, all the elements of row $i$ of matrix A are loaded as a vector into the vector register file. Similarly, all the elements of the corresponding row of matrix $B$ are also loaded as a vector in line 5.  
The vector multiplication between the first element of the loaded row of $A$ and the entire row of matrix $B$ is performed in line 7, which also accumulates the partial results. The row of $A$ is shifted to the right by one element in line 8 to enable the repetition of the above process for the remaining elements of the same row. The final values of all the elements of the corresponding row of the result matrix $C$ are stored back in memory in line 10. The process is repeated until all the rows of matrix $C$ are produced.

\begin{algorithm}[h]
\caption{Row-wise vector matrix multiplication}
\label{a:vectorized_row-wise}
\begin{algorithmic}[1]
\State set vector length
\For{i=0 \textbf{until} i=num\_of\_rows\_of\_A-1}
   \State vload $A[i, :]$ \algorithmiccomment{load the $i$th row of $A$}
   \For{j=0 \textbf{until} j=num\_of\_columns\_of\_A-1}
     \State vload $B[j, :]$ \algorithmiccomment{load the corresponding row of $B$}
     \State s0 = $A[i, 0]$ \algorithmiccomment{move $A[i, 0]$ to a scalar reg}
     \State $C[i, :]$ += s0 * $B[j, :]$ \algorithmiccomment{scalar-vector mul-acc}
     \State $A[i,:] \gg 1$  \algorithmiccomment{vector slide to the right}
   \EndFor
   \State vstore $C[i, :]$ \algorithmiccomment{store to mem the $i$th row of $C$}
\EndFor
\end{algorithmic} 
\end{algorithm}

The baseline vectorized row-based matrix multiplication algorithm can be reformulated to support a \textit{structured-sparse} matrix $A$. This is shown in Algorithm~\ref{a:vectorized_row-wise_sparse-dense}, where all mandatory changes/additions are highlighted with color. The first essential difference pertains to the loading of matrix $A$. Instead of whole rows of $A$, the {\tt values} and {\tt col\_idx} (index) vectors that refer \textit{only} to the \textit{non-zero} elements of $A$ are loaded (lines 3 and 4). The second fundamental difference concerns the selection of the corresponding rows of $B$. In this case, only the rows that refer to the column indexes of the non-zero elements of $A$ are selected (lines 7 and 8) to participate in the multiply-and-accumulate operation (line 10). To move to the next non-zero element, the {\tt values} and {\tt col\_idx} vectors are slid to the right (lines 11 and 12). 

\begin{algorithm}[t]
	\caption{Row-wise vector sparse-dense matrix multiplication}
    \label{a:vectorized_row-wise_sparse-dense}
	\begin{algorithmic}[1]
        \State set vector length
		\For{i=0 \textbf{until} i=num\_of\_rows\_of\_A-1}
              \State \textcolor{olive}{vload values[i, :]} \algorithmiccomment{load $i$th row of values of $A$}
              \State \textcolor{olive}{vload col\_idx[i, :]}  \algorithmiccomment{load ith row of col. indexes of $A$}
              \State \textcolor{olive}{col\_idx[i, :] += B\_address}   \algorithmiccomment{adjust load addresses}
            \For{j=0 \textbf{until} j=\textcolor{olive}{nonzero\_elems\_per\_block}}
                \State \textcolor{olive}{row=col\_idx[i, 0]}
                \State vload $B[\text{\color{olive} row}, :]$               
                \algorithmiccomment{load the selected row of $B$}
                \State s0 = \textcolor{olive}{values[i, 0]}
                \algorithmiccomment{move values[i, 0] to scalar reg}
                \State $C[i, :]$ += s0 * $B[\text{\color{olive} row}, :]$
                \algorithmiccomment{scalar-vector mul-acc}
                \State \textcolor{olive}{values[i, :] $\gg 1$} 
                \algorithmiccomment{vector slide to the right}
                \State \textcolor{olive}{col\_idx[i, :] $\gg 1$} 
                \algorithmiccomment{vector slide to the right}
            \EndFor
            \State vstore $C[i, :]$
            \algorithmiccomment{store to mem the $i$th row of $C$}
		\EndFor
  \end{algorithmic} 
\end{algorithm}

\section{Optimizing sparse-dense matrix multiplication}
\label{s:optimization}

The implementation of sparse-dense multiplication eliminates unnecessary multiplications, due to the structured-sparse format of $A$. However, one crucial bottleneck of the computation is the abundance of \textit{vector loads} from memory for the rows of matrix $B$, as shown in lines 8 and 9 of Algorithm~\ref{a:vectorized_row-wise_sparse-dense}. To tackle this issue, we leverage the structured sparsity of matrix $A$ to reduce memory traffic and allow the computations to use \textit{local} data that already reside in the vector register file.

The proposed optimization combines: (a) the pre-loading of tiles of matrix $B$ in the vector register file, where they are kept stationary for as long as they are needed by the computations; and (b) a custom index-multiply-accumulate instruction that replaces the vector loads of matrix $B$ with lower-cost indirect reads of the vector register file. 

The key attribute that enables the pre-loading of tiles of matrix $B$ in the register file is the well-defined, regular structure in the format of matrix $A$. Since the sparsity of $A$ is -- by construction -- \textit{structured}, the blocks within said matrix have a constant and known size. In turn, this implies that the column indexes of the non-zero element values in $A$ are `bounded' by the block size $M$, i.e., all {\tt col\_idx} values reside within the range $[0, M-1]$. Recall that the block size $M$ is the number of consecutive elements within a row of $A$ that can contain up to a specific number ($N$) of non-zero elements. Exploiting this trait of structured sparsity, we may pre-load as many rows of matrix $B$ as our vector register file can accommodate (with some restrictions, as will be explained shortly) and be sure that the column indexes in matrix $A$ will only point to those local rows. On the contrary, with unstructured sparsity, the column indexes are, essentially, `unbounded' and could point to any row of $B$ (thereby rendering the pre-loading of specific rows of $B$ in the register file futile).

In almost all practical cases, the sizes of matrices $A$ and $B$ are larger than the hardware-supported vector length. Therefore, matrix multiplication is executed in \textit{tiles}. The tile size of matrix $B$ that is pre-loaded in the vector register file is $L\times Vector\_Length$; i.e., $L$ rows, with each row having $Vector\_Length$ columns. The number of rows $L$ must be a multiple of $M$. 
Since a vector register can hold at most $Vector\_Length$ elements of a row of $A$, this means that, for $N$:$M$ structured sparsity, these elements refer to $Vector\_Length/N$ blocks of this row, where each block includes $M$ columns. Thus, the total number of columns of a row of a $A$ and, effectively, the total number of rows of $B$ that can be addressed is \textit{at most} equal to $M\times Vector\_Length/N$. This sets the upper limit on how many rows $L$ of $B$ makes sense to pre-load into the vector register file. 
If more rows of $B$ are pre-loaded, then they would simply not be accessed. 
Of course, pre-loading fewer rows is possible, as long as their number is a multiple of $M$.

\subsection{The proposed index-multiply-accumulate instruction}

The purpose of the vector index-multiply-accumulate ({\tt vindexmac}) instruction is to transform the vector load and multiply-accumulate operations found in lines 8--11 of Algorithm~\ref{a:vectorized_row-wise_sparse-dense} into a new combined operation that would, instead, read the corresponding pre-loaded rows of matrix $B$ directly from the vector register file and operate on them. The {\tt vindexmac} instruction is defined as follows:
{\tt \colorbox{lightgray!40}{vindexmac.vx vd, vs2, rs}}

\begin{figure}[t]
\includegraphics[width=0.98\columnwidth]{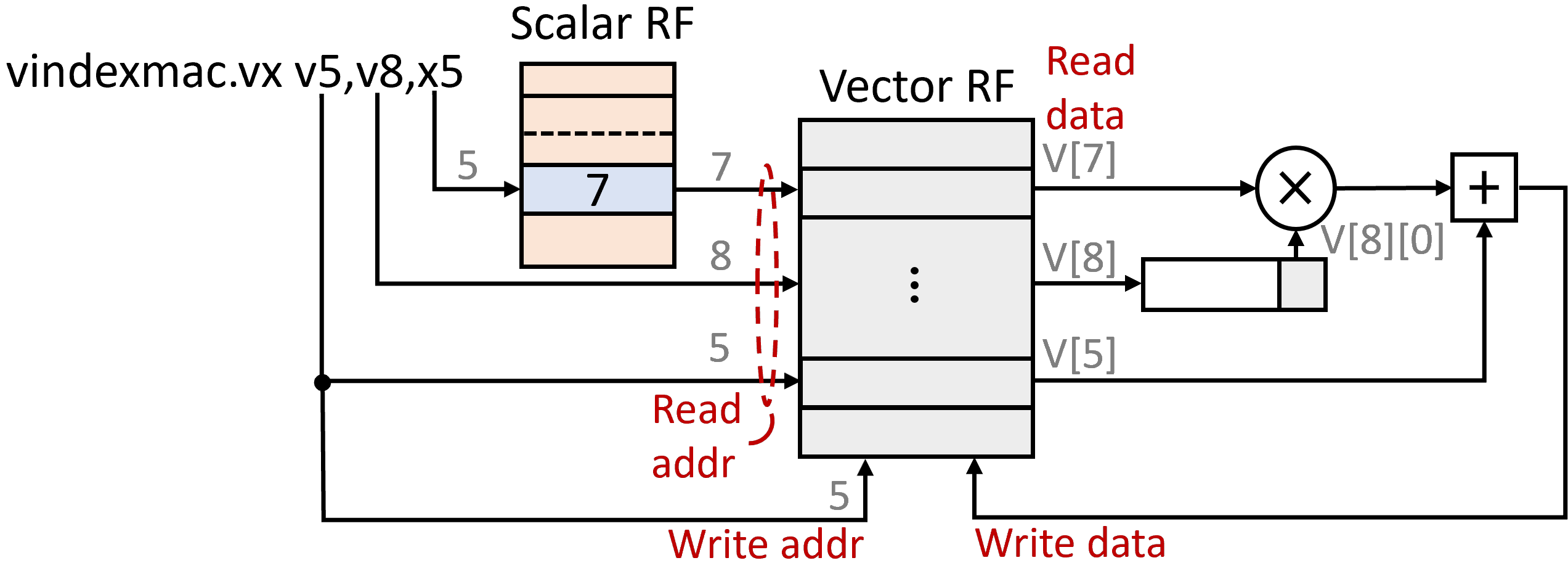}
\caption{The operation of the proposed {\tt vindexmac} instruction. The contents of the scalar register are used to address a specific vector register. The vector read is multiplied with the least significant element of another vector register that is read in parallel. 
The result of the multiplication is accumulated with the previous contents of the vector destination register.}
\label{f:indexmac}
\end{figure}

{\tt vd[i] += vs2[0] * vrf[rs|$_{\texttt{4:0}}$|][i]\\}
\noindent where {\tt vrf} refers to the vector register file. The instruction has three operands: one vector destination register ({\tt vd}) and two source registers. One source register is scalar ({\tt rs}) and the other source register is a vector ({\tt vs2}). To execute {\tt vindexmac}, the scalar register {\tt rs} is read and its contents (only the 5 least significant bits are actually needed) are used as an address to the vector register file. The contents of the vector register that are read via the address contained in {\tt rs} are multiplied with the least significant element of vector register {\tt vs2} and accumulated with the contents of  
{\tt vd}. The operation of {\tt vindexmac} is visualized in Fig.~\ref{f:indexmac}.

The usage of the new {\tt vindexmac} instruction to significantly speed up the sparse-dense matrix multiplication of Algorithm~\ref{a:vectorized_row-wise_sparse-dense} is depicted in Algorithm~\ref{a:vectorized_sparse-dense_indexmac}, \emph{for one tile of matrix $B$}. To complete the matrix multiplication, Algorithm~\ref{a:vectorized_sparse-dense_indexmac} should be \textit{repeated for all tiles}. 
As shown in lines 2--4, a tile of matrix $B$ is pre-loaded into the vector register file. In lines 6 and 7, the non-zero values and column indexes of a row of matrix $A$ are loaded also into the vector register file, similar to Algorithm~\ref{a:vectorized_row-wise_sparse-dense}. The essential differences between Algorithm~\ref{a:vectorized_sparse-dense_indexmac} and the previous algorithms are observed in lines 10 and 11. In line 10, the value that is transferred to the scalar register is the column index needed by {\tt vindexmac} instruction, which is 
executed in line 11 and it replaces the standard scalar-vector multiply-accumulate operation of the previous algorithms. Since the tile of matrix $B$ remains stationary in the vector register file, part of output matrix $C$ is reloaded, updated for each non-zero element of matrix $A$, and stored back to memory in lines 8, 11, and 15, respectively.

\begin{algorithm}[thb]
\caption{Vector sparse-dense matrix multiplication with the use of the new {\tt vindexmac} instruction for \textbf{a tile} of $B$}
    \label{a:vectorized_sparse-dense_indexmac}
	\begin{algorithmic}[1]
        \State set vector length
        \For{k=0 \textbf{until} k=L-1}
          \State {\color{olive} vload $B[k, :]$} \algorithmiccomment{preload L rows of $B$ to vector regs}
        \EndFor
        \For{i=0 \textbf{until} i=num\_of\_rows\_of\_A-1}
            \State vload values[i, :]   \algorithmiccomment{load $i$th row of values of $A$}
            \State vload col\_idx[i, :] \algorithmiccomment{load $i$th row of col. indexes of $A$}
            \State \textcolor{olive}{vload $C[i, :]$} \algorithmiccomment{Load the $i$th row of $C$ from mem}
            \For{j=0 \textbf{until} j=nonzero\_elems\_per\_block}
                \State s0 = col\_idx[i, 0] \algorithmiccomment{move index to scalar reg}
                \State \textcolor{olive}{vindexmac.vx $C[i, :]$, values[i, :], s0}
                \State values[i, :] $\gg 1$  \algorithmiccomment{vector slide to the right}
                \State col\_idx[i, :] $\gg 1$ \algorithmiccomment{vector slide to the right}
            \EndFor
            \State vstore $C[i, :]$ \algorithmiccomment{Store to mem the $i$th row of $C$}
		\EndFor
  \end{algorithmic} 
\end{algorithm}

Overall, the {\tt vindexmac} instruction \textit{replaces} the \textit{three} instructions (one of which is a vector load from memory) of lines 8--10 of Algorithm~\ref{a:vectorized_row-wise_sparse-dense} with the \textit{two} instructions of lines 10 and 11 in Algorithm~\ref{a:vectorized_sparse-dense_indexmac}. The vector load from memory has now been eliminated. The {\tt vindexmac} instruction itself calculates a vector of partial results that correspond to a row of result matrix $C$.

\subsection{Hardware support for {\tt vindexmac} execution}

To execute the {\tt vindexmac} instruction, we need to access both the scalar and the vector register files.  This is an inherent attribute of all RISC-V scalar-vector instructions that have the letter `{\tt x}' in the suffix, e.g., {\tt .vx}, {\tt .vxm}, {\tt .wx}, etc.

In this work, we target vector processors that follow a so called \textit{decoupled} architecture, which consists of a scalar core that is responsible for instruction fetching and orchestration of execution, and a vector engine that executes the vector operations received from the scalar core~\cite{tarantula, xuantie, cornell-vector, ara, riscv-vector, vitruvius}. Since, at any given time, there may be many vector instructions that require the values of scalar registers, the scalar core is responsible to transfer these values to the vector processor, together with the vector instructions themselves. 

Therefore, the value of the scalar register {\tt rs} required by {\tt vindexmac} to address the vector register file is already provided by the scalar core.
The given address drives one of the read ports of the vector register file, which outputs the requested vector operand. Another read port is used to read the elements of {\tt vs2} and the third on reads {\tt vd} as required by all multiply-and-accumulate scalar-vector instructions already present in RISC-V. Therefore, the only hardware requirement of the new {\tt vindexmac} instruction is the addition of a multiplexer in front of the address bus of one of the read ports of the vector register file, which selects between {\tt vs1} of normal vector arithmetic operations, or the 5 least significant bits of {\tt rs} (as required by {\tt vindexmac}). 
In other words, the new instruction by re-using the hardware infrastructure of scalar-vector multiply-add instructions does \textit{not} require an additional read port in the vector register file, but only a 5-bit 2-to-1 multiplexer in front of an existing read port.

Similarly, in fully integrated scalar-vector architectures~\cite{m3, arm-sve}, 
the {\tt vindexmac} instruction would be implemented in exactly the same manner as any of the other {\tt .vx} instructions in a fully integrated scalar-vector setup. The only difference is that the scalar value provided would be used to drive one of the read ports of the vector register file rather than participating directly in computation.

Being a {\tt .vx} instruction, {\tt vindexmac} follows the standard encoding dictated by the RISC-V ISA for scalar-vector instructions~\cite{rvv}.  
However, it terms of the instruction functionality, there is a slight deviation from typical scalar-vector operations. The vector register identifier {\tt vs2} is used only for its least significant element {\tt vs2[0]}, i.e., it plays the role of the scalar value. The actual vector is fetched via an indirect read using the 5 least significant bits of {\tt rs} as an address into the vector register file.

\section{Experimental Evaluation}
\label{s:eval}

The experimental evaluation presented in this section aims to demonstrate the effectiveness of the new {\tt vindexmac} instruction when executing state-of-the-art CNNs such as Resnet50~\cite{resnet}, DenseNet121~\cite{densenet} and InceptionV3~\cite{inception}.
The three CNNs where pruned to 
structured block sparsities of 1:4 and 2:4. This pruning and the appropriate fine-tuning (i.e., re-training) were performed using the TensorFlow library and the ImageNet dataset~\cite{imagenet}. 
Our goal is to quantify the impact of 
transforming vector loads from memory into indirect reads from the vector register file -- through the use of the new instruction -- in terms of achieved speedups and reduction in the total number of memory accesses.

\begin{figure*}
\centering
\includegraphics[width=0.95\textwidth]{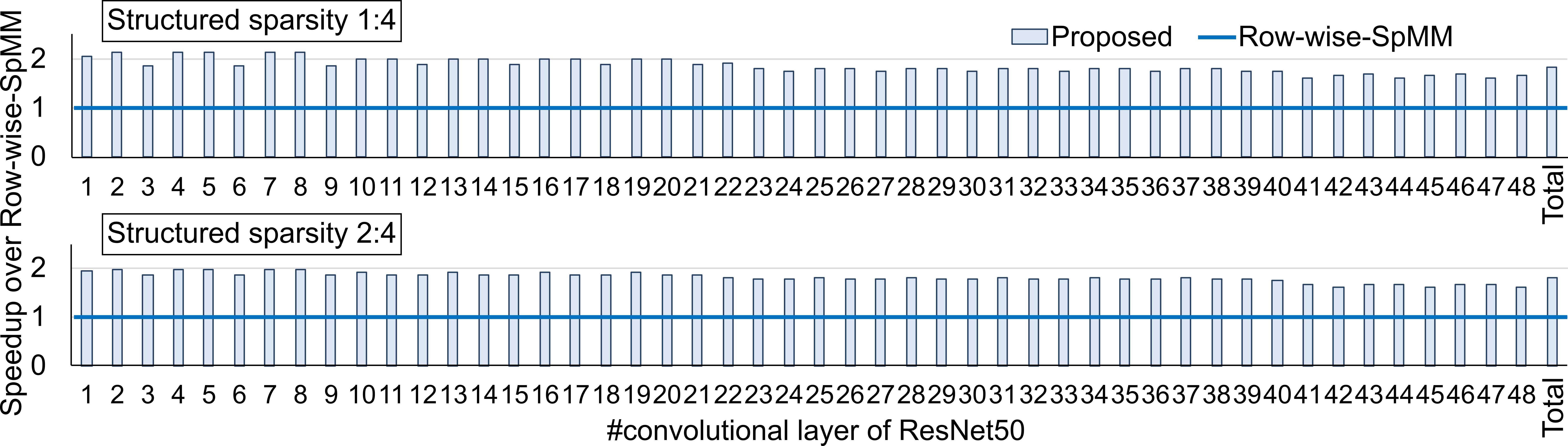}
\caption{The speedup achieved by the proposed approach in the \textit{per-layer} execution times of ResNet50~\cite{resnet}, for (a) 1:4 and (b) 2:4 structured sparsity. The speedup is normalized to the execution time of `Row-Wise-SpMM' in each respective layer. }
\label{f:resnet}
\end{figure*}

\subsection{Experimental setup}
\label{ss:exp-setup}

For all experiments, we utilize a fully implemented decoupled vector unit connected to an out-of-order superscalar processor, i.e., model 1bDV in~\cite{cornell-vector}. This design was modeled in the Gem5 simulator~\cite{gem5-orig, gem5-2020} and the salient design parameters of the simulated processor setup are summarized in Table~\ref{t:proc-det}. The new {\tt vindexmac} instruction was incorporated in the RISC-V GNU toolchain and its operation was implemented in the decoupled vector engine in Gem5.

\begin{table}[t]
\centering
\caption{Simulated Processor Configuration}
\label{t:proc-det}
\renewcommand\tabularxcolumn[1]{m{#1}}
\begin{tabularx}{\columnwidth}{cX}
\hline
Scalar core & 
\begin{itemize}[leftmargin=6pt]
\item 
RISC-V ISA (RV64GC), 8-way-issue out-of-order,
16-entry LSQ, 90 physical integer and 90 physical
 floating-point registers, 60-entry ROB
\item
L1I cache: 1-cycle hit latency, 4-way, 64KB
\item 
L1D cache: 2-cycle hit latency, 4-way, 64KB
\end{itemize}\\
\hline
Vector engine & \begin{itemize}[leftmargin=6pt]
\item 512-bit vector engine with 16-lane configuration (32-bit elements $\times$ 16 execution lanes)
\item The vector engine is connected directly to the L2 cache through 16 store queues and 16 load queues
\end{itemize} \\
\hline
L2 cache & \begin{itemize}[leftmargin=6pt]
\item 8-way, 8-bank
\item 8-cycle hit latency, 512KB shared by both the big core and the vector engine
\end{itemize} \\
\hline
Main Memory & DDR4-2400\\ 
\hline
\end{tabularx}
\end{table}

The convolutions of each layer of the examined CNNs are mapped to sparse-dense matrix multiplications 
$A\times B$~\cite{nvidia-block-sparse}. Matrix $A$ includes the structured-sparse weights and matrix $B$ the input features of the corresponding CNN layers. Input features are considered dense, 
since there is no clear statistical attribute that can be exploited for them. Even if part of the input features contain zero values generated by the corresponding ReLU activation functions in each layer, their number is highly sensitive to the actual input values and filter weights.

The two designs under comparison are (a) {\bf `Row-wise-SpMM'}: the simulated processor setup executing the row-wise sparse-dense matrix multiplications \textit{without} the new instruction, i.e., executing Algorithm~\ref{a:vectorized_row-wise_sparse-dense}; (b) {\bf `Proposed'}: the simulated processor setup executing the row-wise sparse-dense matrix multiplications using Algorithm~\ref{a:vectorized_sparse-dense_indexmac}, i.e., by \textit{employing the new {\tt vindexmac} instruction}.

To increase the performance of both algorithms, we applied loop unrolling, as proposed in~\cite{micro-kernels}, to produce four output results with the corresponding multiply-accumulate instructions in the same loop iteration. Both approaches benefit equally from loop unrolling.

Furthermore, as previously explained, the proposed approach follows a $B$-stationary dataflow to leverage the structured sparsity of $A$ and the new {\tt vindexmac} instruction. In all cases, we assume that the tile of $B$ that is pre-loaded in the vector register file consists of $L$=16 rows. To ensure fairness in the comparisons, we tested \textit{all three} dataflow types for `Row-Wise-SpMM,' i.e., $A$-, $B$-, and $C$-stationary~\cite{micro-kernels}.  
The experimental results show that the $B$-stationary dataflow (used by `Proposed') also yields the best total execution times for `Row-Wise-SpMM.'
Therefore, all the experimental results assume a $B$-stationary dataflow for both approaches under comparison.

\subsection{Evaluation results}

The first set of results, shown in Fig.~\ref{f:resnet}, refer to the execution latencies of each of the CNN layers of ResNet50~\cite{resnet}, for the two examined structured sparsity scenarios (1:4 and 2:4). The obtained execution times are normalized to the performance of `Row-wise-SpMM'. Under 1:4 sparsity, the proposed approach using the new {\tt vindexmac} instruction achieves speedups in the range of 1.60$\times$--2.15$\times$ across all executed CNN layers.
Similarly, under 2:4 sparsity, the obtained speedup is 1.63$\times$--1.99$\times$ across all layers.

Under both sparsities, the speedup tends to slightly decrease as the execution reaches the latter stages of the CNN. As the stages progress, the matrix containing the input features (i.e., matrix $B$) becomes smaller, thus the benefit of pre-loading its rows in the vector register file becomes less pronounced. An interesting observation is that the speedup achieved by the proposed approach is slightly lower (across all layers) in the 2:4 sparsity scenario, as compared to 1:2. This is due to the fact that all the operations for matrix $A$ are now twice as many as before, whereas the operations for matrix $B$ remain the same. Thus, the operations for $A$ now contribute substantially more to the total execution time and decrease the percentage contribution of the operations on $B$, which are the optimization target of {\tt vindexmac}.

Similar behavior is observed in the per-layer execution times of the other two examined CNNs, DenseNet121~\cite{densenet} and InceptionV3~\cite{inception}. Thus, those results are omitted for brevity.

\begin{figure}
\centering
\includegraphics[width=0.95\columnwidth]{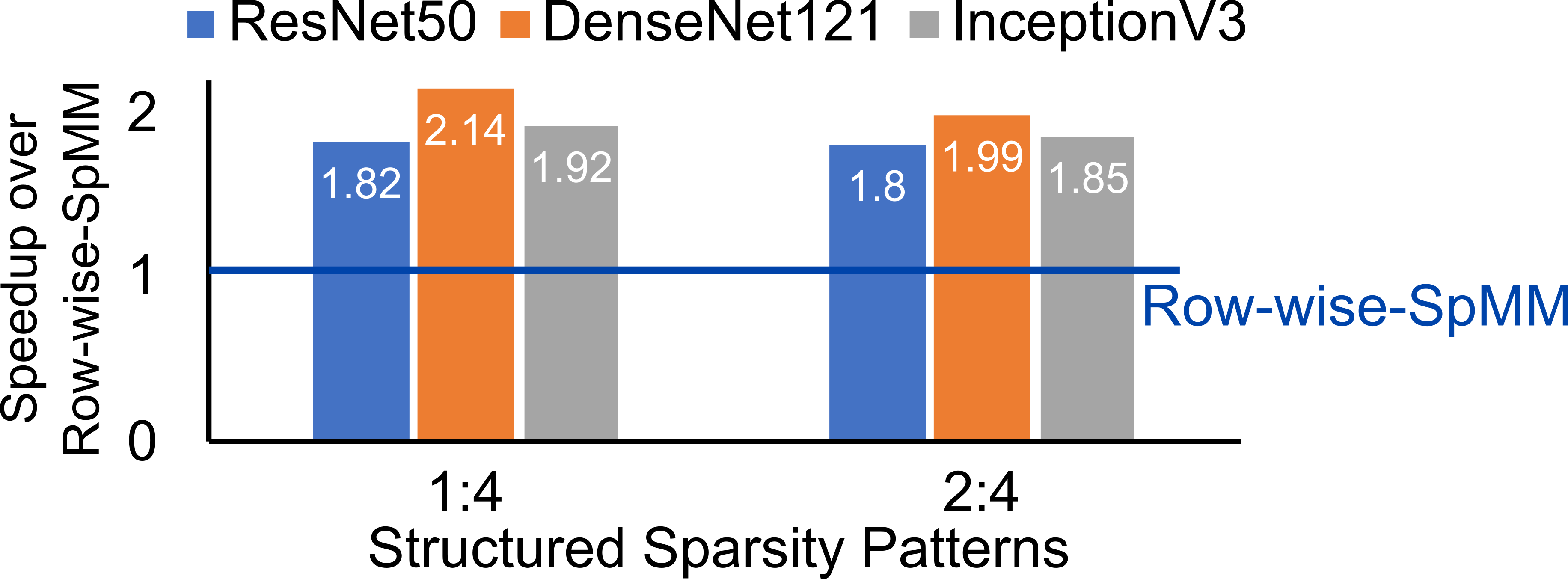}
\caption{The speedup achieved by the proposed approach in the \textit{total} execution times of the three examined CNNs, for (a) 1:4 and (b) 2:4 structured sparsity. The results are normalized to `Row-Wise-SpMM' of the respective sparsity.}
\label{f:total-cnns}
\end{figure}

The speedups achieved in each of the CNN layers translate to significant reductions in the \textit{total} execution times of all three CNNs. Fig.~\ref{f:total-cnns} shows the achieved speedup of the proposed approach over `Row-Wise-SpMM' in the three evaluated CNNs, for 1:4 and 2:4 sparsities. In both sets of bars, the results are normalized to `Row-Wise-SpMM' of the respective sparsity. Clearly, the performance of `Proposed' is substantially better in all cases. Across all three CNNs, the average speedup for 1:4 sparsity is 1.95$\times$, while the average speedup for 2:4 sparsity is 1.88$\times$.

\begin{figure}[t]
    \centering
    \includegraphics[width=0.8\columnwidth]{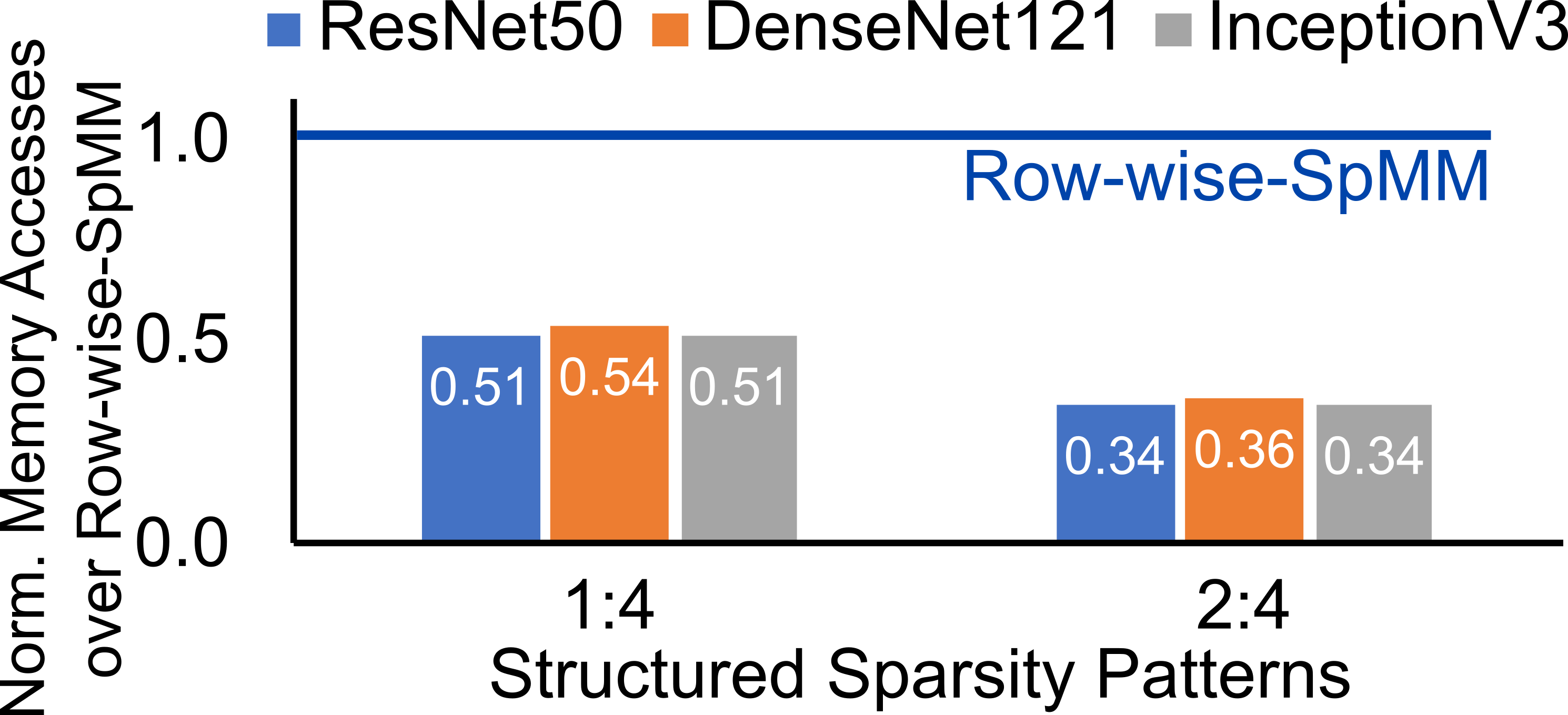}
    \caption{The normalized number of \textit{total} memory accesses observed when using the proposed approach in the three examined CNNs, for (a) 1:4 and (b) 2:4 structured sparsity. The normalization is with respect to `Row-Wise-SpMM' of the respective sparsity.}
    \label{f:vector-load-instructions}
\end{figure}

The above-mentioned improvements in performance when using the `Proposed' approach are reaped from the reduction of vector operations per iteration as well as the 
elimination of unnecessary vector loads from memory and their transformation into indirect reads from the vector register file. By pre-loading tiles of matrix $B$ into the vector register file, the proposed approach exploits data locality very effectively, thereby lowering the memory traffic. The results presented in Fig.~\ref{f:vector-load-instructions} quantify the reduction in total memory accesses when using the proposed {\tt vindexmac} instruction. The presented results are normalized to the number of memory accesses observed with `Row-Wise-SpMM' of the respective sparsity. As can be seen, the total memory accesses decrease markedly. For sparsity 1:4, the memory accesses are reduced by 48\%, on average, while the average reduction for 2:4 sparsity is 65\%. Of course, if `Row-Wise-SpMM' were to employ a $C$-stationary (instead of a $B$-stationary) dataflow, its total number of memory \textit{stores} would decrease significantly. However, as explained at the end of Section~\ref{ss:exp-setup}, this reduction in \textit{store} instructions does not improve the total execution time.

\section{Conclusions}
\label{s:conclusions}

Vector processors can efficiently handle the abundant data-level parallelism available in modern ML applications. The scalability of ML models calls for appropriate model pruning that reduces their memory footprint and makes them amenable to applications at the edge. In this context, we aimed to seamlessly integrate the simplicity of structured sparsity with mainstream vector architectures enhanced with a proposed new instruction. The {\tt vindexmac} instruction operates on local data that is pre-loaded into the vector register file, thereby reducing the number of instructions executed and eliminating unnecessary vector loads.  
Most importantly, the new instruction can be implemented with negligible hardware cost. The evaluation results demonstrate substantial speedups in the execution latency of representative layers of state-of-the-art CNN applications.

\bibliographystyle{IEEEtran}
\bibliography{refs}
\end{document}